\documentclass[twoside,twocolumn,9pt]{article}
\usepackage{extsizes}
\usepackage[super,sort&compress,comma]{natbib} 
\usepackage[version=3]{mhchem}
\usepackage[left=1.5cm, right=1.5cm, top=1.785cm, bottom=2.0cm]{geometry}
\usepackage{balance}
\usepackage{mathptmx}
\usepackage{sectsty}
\usepackage{graphicx} 
\usepackage{lastpage}
\usepackage[format=plain,justification=justified,singlelinecheck=false,font={stretch=1.125,small,sf},labelfont=bf,labelsep=space]{caption}
\usepackage{float}
\usepackage{fancyhdr}
\usepackage{fnpos}
\usepackage[english]{babel}
\addto{\captionsenglish}{%
  \renewcommand{\refname}{Notes and references}
}
\usepackage{array}
\usepackage{droidsans}
\usepackage{charter}
\usepackage[T1]{fontenc}
\usepackage[usenames,dvipsnames]{xcolor}
\usepackage{setspace}
\usepackage[compact]{titlesec}
\usepackage{hyperref}

\usepackage{epstopdf}

\definecolor{cream}{RGB}{222,217,201}

\begin{document}

\pagestyle{fancy}
\thispagestyle{plain}
\fancypagestyle{plain}{
\renewcommand{\headrulewidth}{0pt}
}

\makeFNbottom
\makeatletter
\renewcommand\LARGE{\@setfontsize\LARGE{15pt}{17}}
\renewcommand\Large{\@setfontsize\Large{12pt}{14}}
\renewcommand\large{\@setfontsize\large{10pt}{12}}
\renewcommand\footnotesize{\@setfontsize\footnotesize{7pt}{10}}
\makeatother

\renewcommand{\thefootnote}{\fnsymbol{footnote}}
\renewcommand\footnoterule{\vspace*{1pt}%
\color{cream}\hrule width 3.5in height 0.4pt \color{black}\vspace*{5pt}} 
\setcounter{secnumdepth}{5}

\makeatletter 
\renewcommand\@biblabel[1]{#1}            
\renewcommand\@makefntext[1]%
{\noindent\makebox[0pt][r]{\@thefnmark\,}#1}
\makeatother 
\renewcommand{\figurename}{\small{Fig.}~}
\sectionfont{\sffamily\Large}
\subsectionfont{\normalsize}
\subsubsectionfont{\bf}
\setstretch{1.125} 
\setlength{\skip\footins}{0.8cm}
\setlength{\footnotesep}{0.25cm}
\setlength{\jot}{10pt}
\titlespacing*{\section}{0pt}{4pt}{4pt}
\titlespacing*{\subsection}{0pt}{15pt}{1pt}

\fancyfoot{}
\fancyfoot[LO,RE]{\vspace{-7.1pt}\includegraphics[height=9pt]{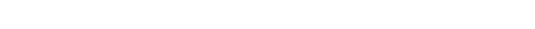}}
\fancyfoot[CO]{\vspace{-7.1pt}\hspace{11.9cm}\includegraphics{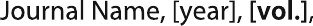}}
\fancyfoot[CE]{\vspace{-7.2pt}\hspace{-13.2cm}\includegraphics{head_foot/RF}}
\fancyfoot[RO]{\footnotesize{\sffamily{1--\pageref{LastPage} ~\textbar  \hspace{2pt}\thepage}}}
\fancyfoot[LE]{\footnotesize{\sffamily{\thepage~\textbar\hspace{4.65cm} 1--\pageref{LastPage}}}}
\fancyhead{}
\renewcommand{\headrulewidth}{0pt} 
\renewcommand{\footrulewidth}{0pt}
\setlength{\arrayrulewidth}{1pt}
\setlength{\columnsep}{6.5mm}
\setlength\bibsep{1pt}

\makeatletter 
\newlength{\figrulesep} 
\setlength{\figrulesep}{0.5\textfloatsep} 

\newcommand{\topfigrule}{\vspace*{-1pt}%
\noindent{\color{cream}\rule[-\figrulesep]{\columnwidth}{1.5pt}} }

\newcommand{\botfigrule}{\vspace*{-2pt}%
\noindent{\color{cream}\rule[\figrulesep]{\columnwidth}{1.5pt}} }

\newcommand{\dblfigrule}{\vspace*{-1pt}%
\noindent{\color{cream}\rule[-\figrulesep]{\textwidth}{1.5pt}} }

\makeatother

\twocolumn[
  \begin{@twocolumnfalse}
{\includegraphics[height=30pt]{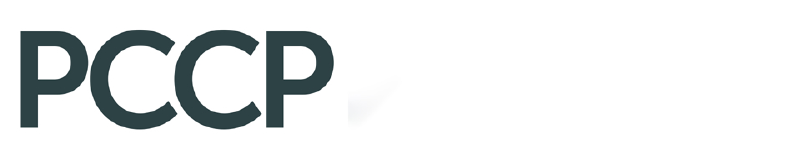}\hfill\raisebox{0pt}[0pt][0pt]{\includegraphics[height=55pt]{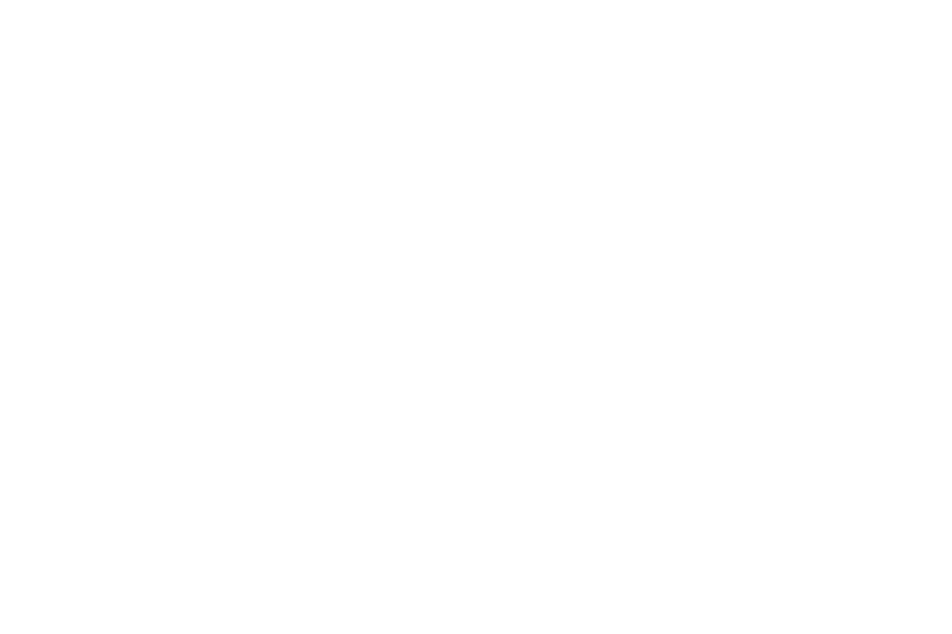}}\\[1ex]
\includegraphics[width=18.5cm]{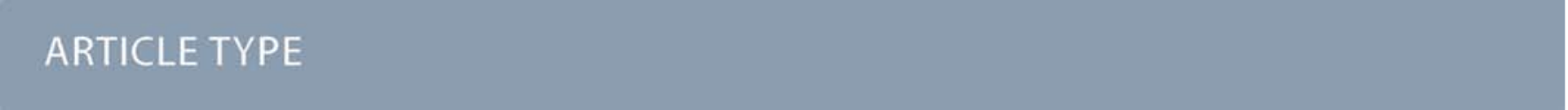}}\par
\vspace{1em}
\sffamily
\begin{tabular}{m{4.5cm} p{13.5cm} }

\includegraphics{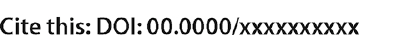} & \noindent\LARGE{\textbf{Secondary ionization of pyrimidine nucleobases and their microhydrated derivatives in helium nanodroplets}} \\
\vspace{0.3cm} & \vspace{0.3cm} \\

 & \noindent\large{Jakob D. Asmussen,\textit{$^{a}$} Abdul R. Abid,\textit{$^{a}$} Akgash Sundaralingam,\textit{$^{a}$} Björn Bastian,\textit{$^{a}$} Keshav Sishodia,\textit{$^{b}$} Subhendu De,\textit{$^{b}$} Ltaief Ben Ltaief,\textit{$^{a}$} Sivarama R. Krishnan,\textit{$^{b}$} Henrik B. Pedersen,\textit{$^{a}$} and Marcel Mudrich\textit{$^{a,\dag}$}} \\

\includegraphics{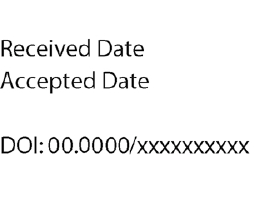} & \noindent\normalsize{Radiation damage in biological systems by ionizing radiation is predominantly caused by secondary processes such as charge and energy transfer leading to the breaking of bonds in DNA. Here, we study the fragmentation of cytosine (Cyt) and thymine (Thy) molecules, clusters and microhydrated derivatives induced by direct and indirect ionization initiated by extreme-ultraviolet (XUV) irradiation. Photofragmentation mass spectra and photoelectron spectra of free Cyt and Thy molecules are compared with mass and electron spectra of Cyt/Thy clusters and microhydrated Cyt/Thy molecules formed by aggregation in superfluid helium (He) nanodroplets. Penning ionization after resonant excitation of the He droplets is generally found to cause less fragmentation compared to direct photoionization and charge-transfer ionization after photoionization of the He droplets. When Cyt/Thy molecules and oligomers are complexed with water molecules, their fragmentation is efficiently suppressed. However, a similar suppression of fragmentation is observed when homogeneous Cyt/Thy clusters are formed in He nanodroplets, indicating a general trend. Penning ionization electron spectra (PIES) of Cyt/Thy are broad and nearly featureless but PIES of their microhydrated derivatives point at a sequential ionization process ending in unfragmented microsolvated Cyt/Thy cations.} \\

\end{tabular}

 \end{@twocolumnfalse} \vspace{0.6cm}

  ]

\renewcommand*\rmdefault{bch}\normalfont\upshape
\rmfamily
\section*{}
\vspace{-1cm}


\footnotetext{\textit{$^{a}$~Department of Physics and Astronomy, Aarhus University, 8000 Aarhus C, Denmark.}}
\footnotetext{\textit{$^{b}$~Quantum Center of Excellence for Diamond and Emergent Materials and Department of Physics, Indian Institute of Technology Madras, Chennai 600036, India. }}
\footnotetext{\textit{$^{\dag}$~E-mail: mudrich@phys.au.dk }}




\section{Introduction}
Ionization of deoxyribonucleic acid (DNA) bases is a key step in radiation damage leading to mutation.~\cite{steenken1989purine} Radiation damage is not only caused by direct impact of high-energy photons on the nucleobases, but to a large extent by secondary particles (electrons, ions and radicals) formed from reactions induced by the ionizing radiation.~\cite{swiderek2006fundamental} The fragmentation of DNA bases upon ionization by collisions with electrons~\cite{sanche2005low} and ions~\cite{rodgers1994low} has extensively been studied to unravel the key processes leading to radiation damage. An important element in understanding the reaction paths to damage of DNA is the interaction of the nucleobases with the aqueous medium which affects the ionization potential~\cite{cauet2010vertical} and can alter the fragmentation pathways due to energy~\cite{kocisek2016microhydration} or proton~\cite{khistyaev2013proton,jacquemin2014assessing,semmeq2020dna} transfer.   

\begin{figure}[t]
    \centering
    \includegraphics[width=0.7\columnwidth]{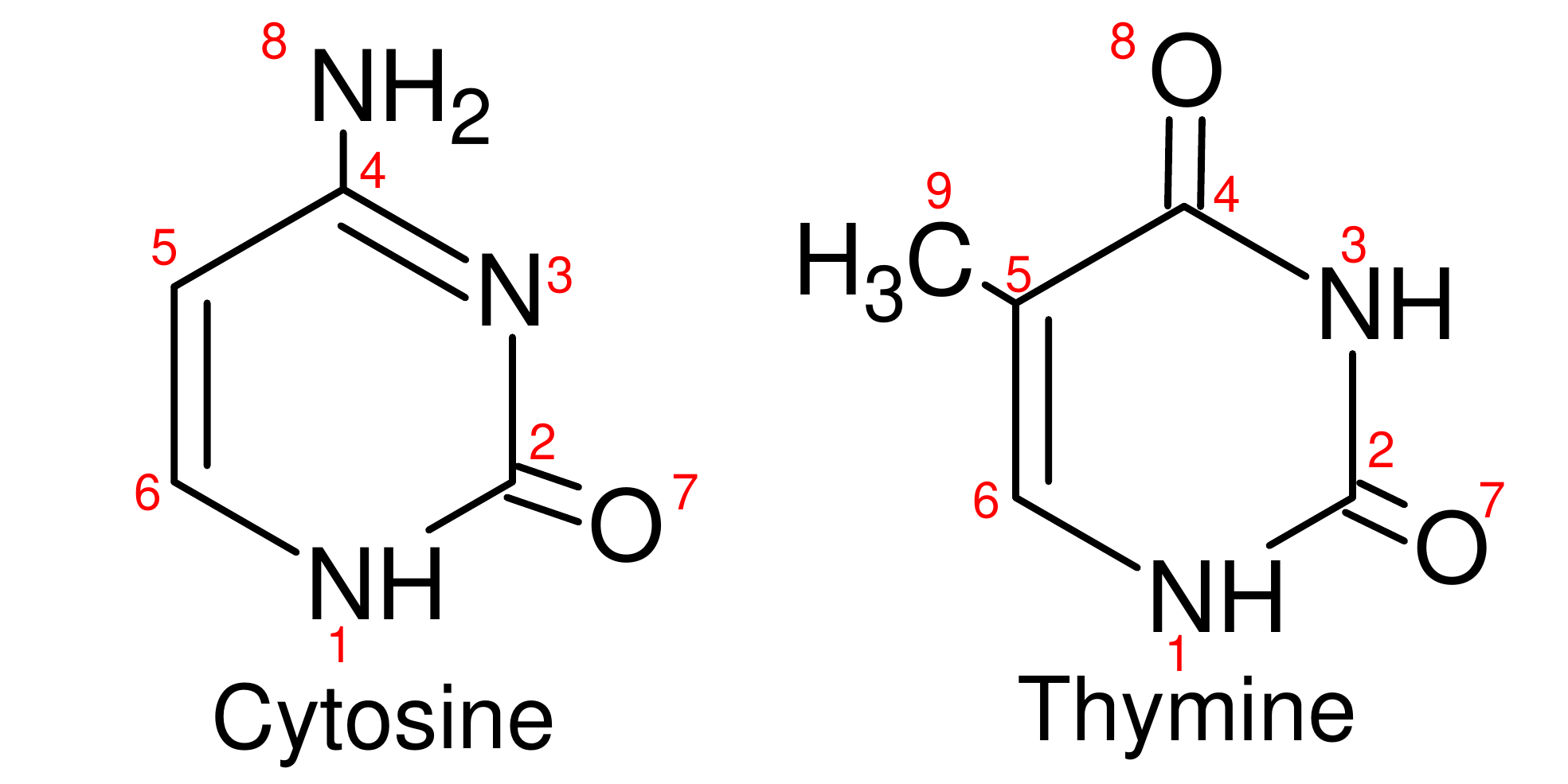}
    \caption{Molecular structure of cytosine and thymine.}
    \label{fig:molecules}
\end{figure}

\begin{figure*}[t]
    \centering
    \includegraphics[width=1.6\columnwidth]{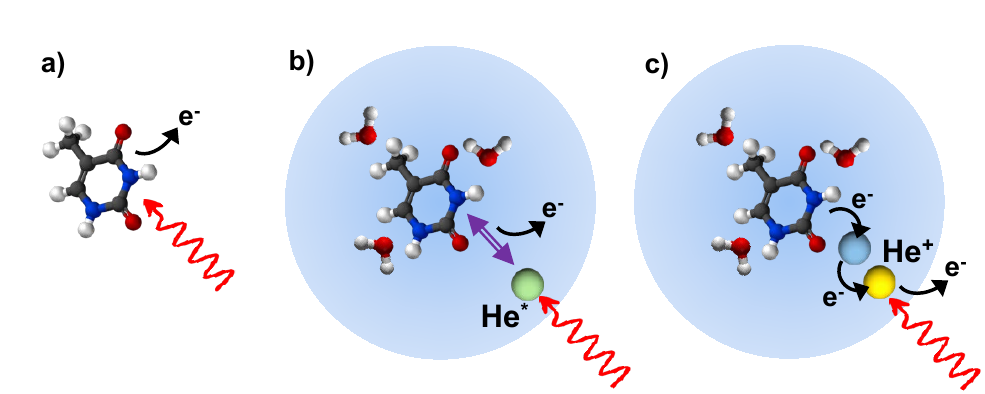}
    \caption{Illustration of the three different ionization schemes for thymine and microhydrated thymine: a) Direct photoionization of the molecule, b) Penning ionization in a helium nanodroplet following resonant excitation ($h\nu = 21.6$~eV) of the droplet and c) charge-transfer ionization in a helium nanodroplet after direct ionization of the droplet ($h\nu = 26$~eV).
    }
    \label{fig:cartoon}
\end{figure*}

Fragmentation of molecular ions is readily studied in the gas phase using molecular beams techniques. Formation of microhydrated clusters (nucleobases weakly bound to a few water molecules) is possible in molecular beams, but controlling the cluster size is difficult.~\cite{kim1996cluster} In this study, we investigate the fragmentation of the two pyrimidine nucleobases cytosine (Cyt) and thymine (Thy) found in DNA  (see Fig.~\ref{fig:molecules}) induced by direct and indirect ionization after XUV irradiation. Free Cyt and Thy molecules in an effusive beam are directly photoionized and Cyt/Thy clusters and microhydrated complexes are formed by means of aggregation in helium (He) nanodroplets (HNDs). HNDs are cold (0.4 K), superfluid clusters of weakly-bound He.~\cite{toennies2004superfluid} Due to their capability to efficiently pick up foreign species ('dopants'), the high mobility of dopants inside HNDs, and their chemical inertness, various clusters can be formed and studied in HNDs. The size of the dopants' clusters is controlled by the partial pressure of the dopant vapor in the pick-up cell and can often be determined from a Poisson distribution.~\cite{toennies1998spectroscopy} In this way we can achieve a relatively high degree of control of the composition of homogeneous and of heterogeneous clusters, such as microhydrated biomolecules. Dopants in HNDs can efficiently be Penning ionized through resonant excitation of the droplet, or through direct ionization of the droplet leading to radiative charge-transfer (RCT) ionization of the dopant.~\cite{buchta2013charge,kim2006photoionization} Resonant photoexcitation is most efficient for the HND state excited at a photon energy $h\nu = 21.6$~eV which correlates to the  1s2p\,$^1$P atomic He state.~\cite{joppien1993electronic}  This state relaxes to the metastable 1s2s\,$^1$S He atomic state within about 1~ps.~\cite{mudrich2020ultrafast,laforge2022relaxation} The metastable He atom further Penning ionizes the dopant by decaying to the ground state thereby releasing its energy to the dopant which in turn is ionized.\cite{ben2019charge} 
Penning ionization and RCT ionization are instances of indirect ionization processes in heterogeneous systems induced by energetic radiation, which have analogues in aqueous systems such as biological tissue. ~\cite{Alizadeh:2015,Ren:2018} 

In this study, we aim at characterizing the fragmentation of Cyt and Thy by comparing direct photoionization of the free molecules in the gas-phase with indirect ionization of the molecules embedded in HNDs (see Fig.~\ref{fig:cartoon}). By adjusting the doping level and by co-doping with water molecules, we measure fragment distributions from ionization of Thy and Cyt clusters and of the microhydrated Thy, Cyt molecules. As a general trend, fragmentation is efficiently suppressed when the molecules are complexed in clusters. From Penning ionization electron spectra (PIES) recorded in coincidence with various Thy, Cyt fragments and their complexes with water, we obtain some insight into the ionization process. Based on these results, we assess the benefit of HNDs as nano-matrices for studies of photoionization and fragmentation processes with relevance to radiation biology.

\section{Methods}
Ion mass spectra and electron velocity map images (VMIs) were measured using the XENIA (XUV electron-ion spectrometer for nanodroplets in Aarhus) endstation~\cite{bastian2022new} located at the AMOLine of the ASTRID2 synchrotron at Aarhus University, Denmark.~\cite{hertel2011astrid2} Using photoelectron-photoion coincidence (PEPICO) detection, VMIs were recorded in coincidence with specific fragment ions of Thy, Cyt molecules and clusters. Ion and electron yields were background-subtracted using a rotating chopper which periodically blocks and unblocks the HND beam. At the photon energy $h\nu = 21.6$~eV (resonant excitation of HNDs), a tin (Sn) filter was used to block higher harmonics of the undulator radiation; at $h\nu = 26.0$~eV (photoionization of HNDs), an aluminum (Al) filter was used. The photon flux is estimated from the yield of electron-H$_2$O$^+$ coincidences detected from the background gas.~\cite{haddad1986total} Electron spectra were inferred from the VMI by Abel inversion using the MEVELER reconstruction method~\cite{dick2014inverting}.

HNDs were formed by continuous expansion of He at high pressure (30~bar) into vacuum through a cryogenically cooled (14 K) nozzle of diameter $5~\mu$m. The average droplet size is determined from titration measurements to be $\langle N \rangle = 1.9\times10^4$.~\cite{gomez2011sizes} 
The droplets were first doped with Cyt or Thy by passing them through a 1~cm long vapor cell. The cell was heated to 140-165\,$^{\circ}$C and 92-108\,$^{\circ}$C for doping with Cyt and Thy, respectively. The doping level for the two dopants is mainly determined from the monomer-to-dimer ratio detected at $h\nu = 21.6$~eV for an oven temperature of 140\,$^{\circ}$C and 90\,$^{\circ}$C, respectively. Changes of the doping level from this value is then determined based on the change in vapor pressure as a function of varying oven temperature.~\cite{ferro1980vapour}     
Subsequently, the HNDs were doped with H$_2$O or D$_2$O by leaking water vapor into a gas doping cell of length 1.8~cm further downstream. The doping level was determined using the formula derived by Kuma \textit{et al.}~\cite{kuma2007laser}
An effusive molecular beam of Cyt or Thy was realized by heating an effusive cell with a nozzle opening of 1~mm diameter. The cell was heated to 220$^{\circ}$\,C to create an effusive beam of Cyt and to 170$^{\circ}$\,C to create an effusive beam of Thy. 

\section{Results and discussion}
\begin{figure*}[t]
    \centering
    \includegraphics[width=1.9\columnwidth]{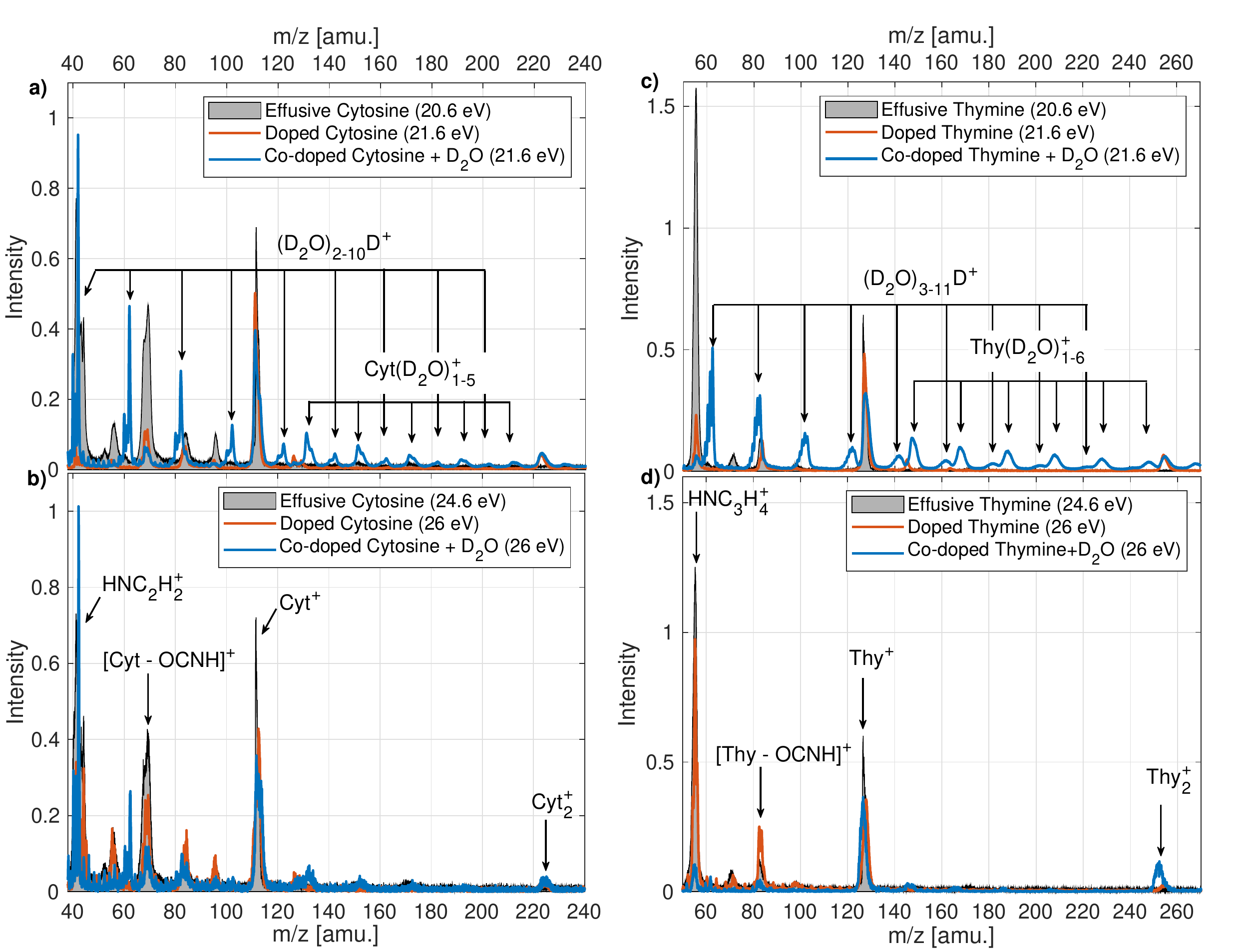}
    \caption{Mass spectra recorded for Cyt [a) \& b)] and Thy [c) \& d)] in an effusive molecular beam, doped into HNDs or co-doped with D$_2$O into HNDs. a) \& c) show the mass spectra for Penning ionization of the dopant ($h\nu = 21.6$~eV), and b) \& d) show mass spectra for charge-transfer ionization ($h\nu = 26$~eV). In case of photoionization of the effusive beam, the photon energy is matched to the internal energy of the excited He atom inducing Penning ionization and to the He ion inducing charge-transfer ionization.    }
    \label{fig:mass_spectra}
\end{figure*}

\begin{figure*}[t]
    \centering
    \includegraphics[width=1.4\columnwidth]{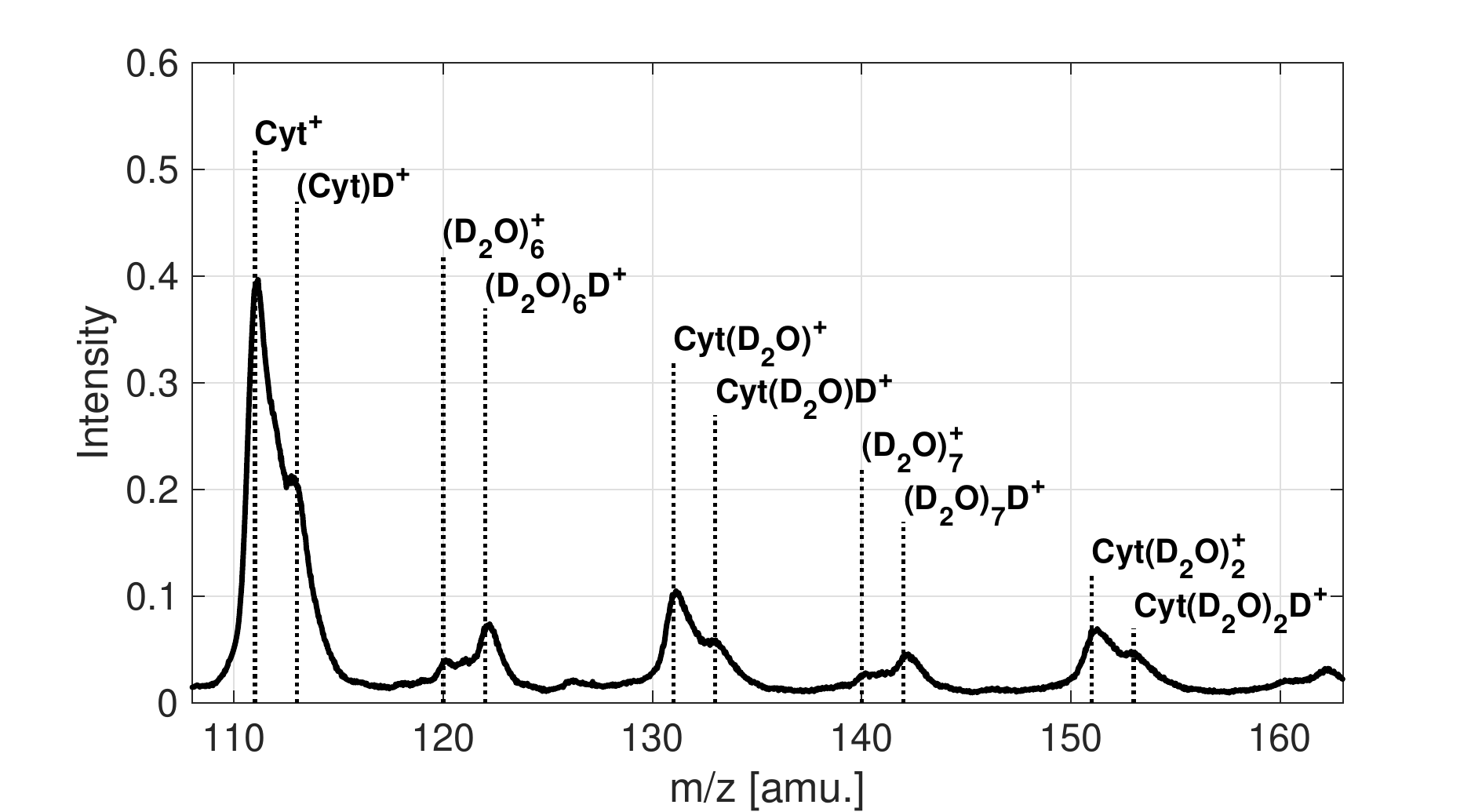}
    \caption{Penning ionization mass spectrum ($h\nu = 21.6$~eV) of microhydrated Cyt in He nanodroplets.  }
    \label{fig:cyt_zoomin}
\end{figure*}
\subsection{Ion mass spectra}
In Fig.~\ref{fig:mass_spectra}, we present the ion mass spectra recorded by either photoionizing free Cyt and Thy molecules in an effusive beam or by indirectly ionizing Cyt and Thy embedded in HNDs at two photon energies $h\nu = 21.6$~eV (Penning ionization) and $h\nu = 26$~eV (RCT ionization). Additionally, mass spectra of microhydrated Cyt and Thy formed by co-doping the HNDs with Cyt/Thy and with D$_2$O molecules are shown (blue lines). In the case of photoionization of the effusive beam, the photon energy is tuned to 20.6~eV and 24.6~eV; these values match the internal energy of the metastable He atom and the He ion which is released to the Cyt/Thy dopants by Penning ionization, respectively RCT ionization processes. 

The various fragmentation channels of Cyt~\cite{plekan2007photofragmentation,sadr2015fragmentation,shafranyosh2022electronic} and Thy~\cite{imhoff2005identification,van2014electron,majer2019valence} have been identified by means of high-resolution mass spectrometry and quantum chemical calculations. From photoionization of effusive Cyt and Thy, we identify two main fragments in agreement with mass spectra reported in the literature. The main fragmentation pathway of the parent cation involves the loss of isocyanic acid (OCNH) through a retro-Diels-Alder (rDA) reaction giving rise to the $m/z = 68$ and $m/z=83$ for Cyt and Thy, respectively. The rDA reaction happens through breaking of the bonds N1-C2 and N3-C4 (see Fig.~\ref{fig:molecules}).~\cite{sadr2015fragmentation,jochims2005photoion} Both fragments are subjected to further fragmentation. The product of the rDA reaction of Cyt further fragments through the elimination of HCN (or NHC) as a result of break-up of the C5-C6 bond (or C4-C5).~\cite{sadr2015fragmentation} This leads to the HNC$_2$H$_2^+$ fragment ($m/z = 41$). The Thy rDA product fragments through elimination of either CO or HCN (breaking of C4-C5 or C5-C6 bonds) resulting in fragments of masses $m/z=55$ and $m/z=56$.~\cite{jochims2005photoion} Due to our limited mass resolution, we cannot fully separate the two ionic fragments in the effusive beam mass spectrum, but based on the center of joint mass peak, we identify CO elimination as the main pathway in consistence with previous findings.~\cite{imhoff2005identification} Accordingly, the peak is identified as HNC$_3$H$_4^+$ ($m/z=55$). In general, minor fragmentation paths leading to the aforementioned fragments with one or two missing or added hydrogen atoms are expected to be present but cannot be resolved with our spectrometer. 

Indirect ionization of Cyt/Thy embedded in HNDs leads to similar fragments as direct photoionization of free Cyt/Thy. Additionally, we detect Cyt$_2^+$, Thy$_2^+$ clusters. We do not detect any fragments with masses between the monomer and dimer parent ions indicating that fragmentation of clusters in HNDs is limited to loss of intact parent moieties. 
In the case of RCT ionization ($h\nu = 26$~eV), the relative yield of the fragments to the parent ion only slightly differs from that detected by photoionization of the effusive beam, whereas in the case of Penning ionization ($h\nu = 21.6$~eV) the yield of the fragments ions are significantly suppressed. This shows that Penning ionization is a ``softer'' ionization channel where the excess energy is dissipated in the droplet thereby suppressing fragmentation. Previous studies of electron-impact ionization of doped HNDs have reported reduced fragmentation of dopants in the droplet.~\cite{lewis2004electron,yang2006electron,denifl2009electron} 
One study found that in particular fragmentation involving the break-up of C-C bonds was suppressed in HNDs~\cite{yang2006electron} which matches well the observed reduction of the yields of fragments formed by HCN, NHC and CO elimination (relying on break-up of the C-C bonds) and increased yields of fragments from rDA (relying on break-up of N-C bonds).

In the case of microhydration of Cyt/Thy by co-doping with  D$_2$O, the fragmentation of the pyrimidines is further suppressed. For Penning ionization, a series of D$_2$O cluster ions and Cyt/Thy ions with attached D$_2$O molecules can be seen in the mass spectra (see Fig.~\ref{fig:cyt_zoomin} for Penning ionization of microhydrated cytosine), whereas in the case of RCT ionization these cluster ions are less prominent. This further confirms the fact that Penning ionization is a softer ionization channel compared to RCT ionization. 
For the water cluster ions, the deuterated cluster, (D$_2$O)$_n$D$^+$, dominates over the undeterated cluster, (D$_2$O)$_n^+$. Previous reports show that the unprotonated cluster is not observed for ionization of bare water clusters.\cite{shiromaru1987synchrotron,jongma1998rapid} However, in HNDs the ejection of the OH, OD radical following proton-, respectively deuteron transfer can be inhibited.\cite{yang2007electron,denifl2009ion} For the microhydrated Cyt/Thy ion complex, the undeuterated cluster dominates over the deuterated cluster. Due to the high proton-affinity of the nucleobases,~\cite{meot1979ion} one would expect the deuteron-transfer to be efficient towards Cyt/Thy. The fact that the undeuterated ion complexes have higher yields suggest that the ejection of the OH (OD) radical is less efficient for the mixed cluster. Denifl \emph{et al.} have reported similar findings for ionization of HNDs co-doped with fullerene and water.\cite{denifl2009ion}  



\begin{figure}[t]
    \centering
    \includegraphics[width=0.8\columnwidth]{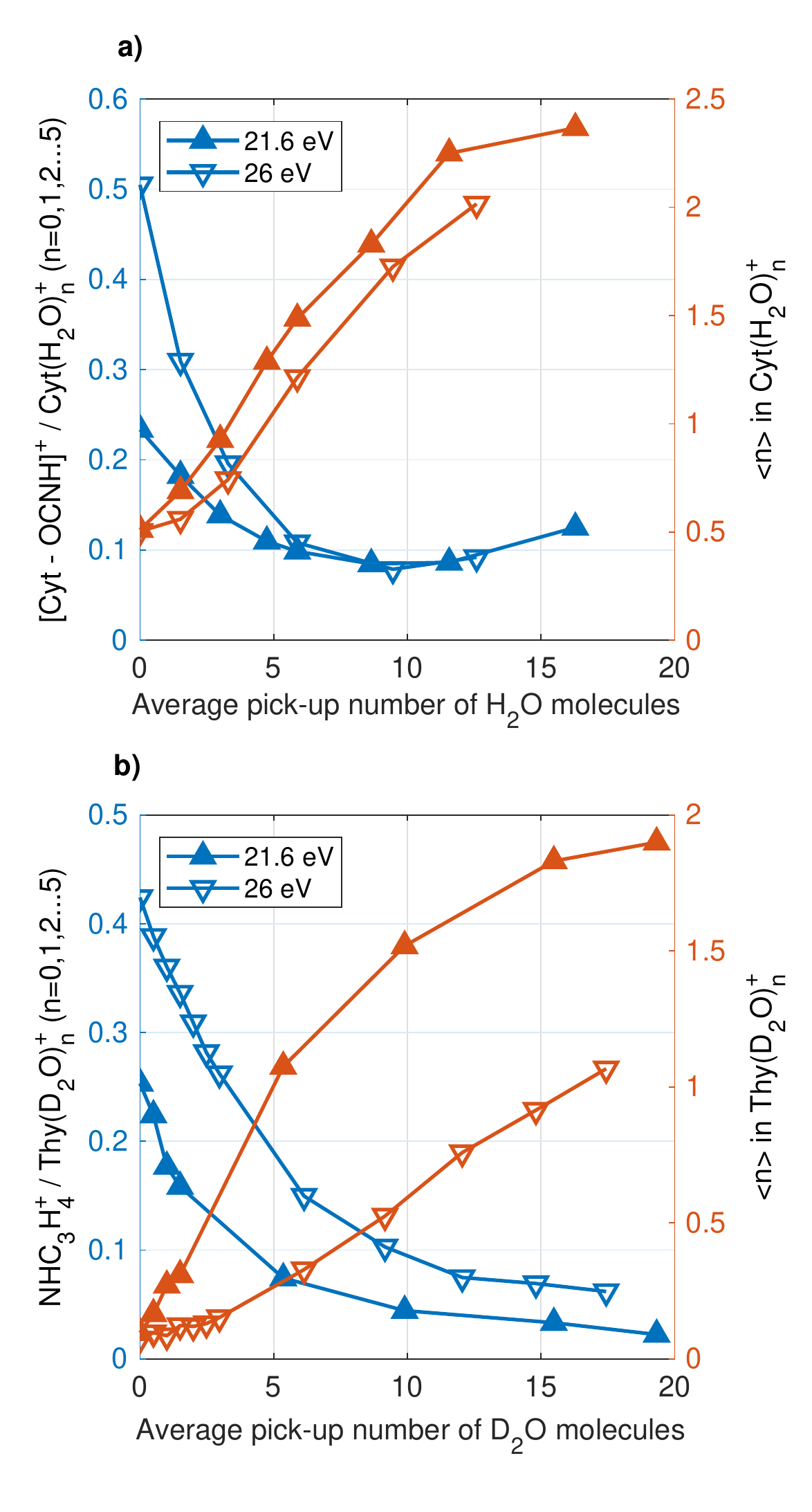}
    \caption{Relative yields of [Cyt-OCNH]$^+$ to Cyt(H$_2$O)$_{0-5}^+$ (a) and of NHC$_3$H$_4^+$ to Thy(D$_2$O)$_{0-5}^+$ (b) as a function of H$_2$O/D$_2$O doping level (blue symbols, left axis). The red symbols and right axes show the average number of water molecules bound to the Cyt/Thy parent ions. For both dopants, the relative fragment yield and average number of attached water molecules are shown for Penning ionization ($h\nu = 21.6$~eV) and charge-transfer ionization ($h\nu = 26$~eV). }
    \label{fig:frag_codoped}
\end{figure}

Fig.~\ref{fig:frag_codoped}~(red symbols, right axis) shows the average size, $\langle n \rangle$, of the Cyt(H$_2$O)$_n^+$, Thy(D$_2$O)$_n^+$ ions detected as function of water doping level. 
Note that in the case of Cyt, we have replaced D$_2$O as co-dopant by H$_2$O as there was no particular advantages of using deuterated water. However, use of D$_2$O avoids the overlap in mass between the Thy parent ion and the (H$_2$O)$_7^+$ cluster. 
The average number of water molecules bound to the ionic complex rises with increasing doping level, but does not reflect the actual number of water molecules surrounding the Cyt/Thy molecule in the  droplet since the average number of water molecules in the ionic complex is smaller than the average number of doped water molecules. We note that larger clusters ($n\geq6$) may be present in mass spectra but are not considered here due to the limited resolution for larger masses. Thus, $\langle n \rangle$ shown in Fig.~\ref{fig:frag_codoped} is most likely underestimated; however it is unlikely that this explains the large difference (nearly factor 10) between the detected ion cluster size and the actual dopant cluster size. Thus, the microhydrated Cyt/Thy clusters fragment into smaller clusters (elimination of [H$_2$O]$_n$/[D$_2$O]$_n$) upon Penning and RCT ionization, where fragmentation is enhanced for RCT ionization as compared to Penning ionization.
The rise of $\langle n \rangle$ seems to level out for high doping strength possibly indicating the presence of a hydration shell around the Cyt/Thy ion. A study of microhydrated Thy (and adenine) has shown a hydration shell of $n=4$ around the nucleobase ion.~\cite{kim2002hydration} It would require higher doping levels to investigate whether convergence at $n=4$ for the dopant ion is the case. However, we note that the $n=4$ hydrated ion does not not show an enhanced abundance compared to $n=3$ and $n=5$ in the mass spectra [Fig.~\ref{fig:mass_spectra}~a)+c)]. 

A clear trend is that the fragmentation of the Cyt/Thy molecules is suppressed by increasing the level of microhydration. Fig.~\ref{fig:frag_codoped}~(blue symbols, left axis) shows the ratio of fragment ion [Cyt$-$OCNH]$^+$ (NHC$_3$H$_4^+$) to unfragmented ion Cyt(H$_2$O)$_{0-5}^+$ (Thy[D$_2$O]$_{0-5}^+$) yields for increasing doping level of water. We refer to the rDA product [Cyt$-$OCNH]$^+$ for Cyt and the product following CO elimination, NHC$_3$H$_4^+$, for Thy since the respective other main ion fragment of the two molecules overlap with a water cluster ion. Both for Penning ionization and RCT ionization, the ratio of fragment ions to hydrated and non-hydrated parent ion drops when increasing the number of added water molecules. In the case of Cyt, convergence of the ratio at 0.1 for both Penning and RCT ionization reflects the detection limit due to noise in the spectra. 


\begin{figure}[t]
    \centering
    \includegraphics[width=0.8\columnwidth]{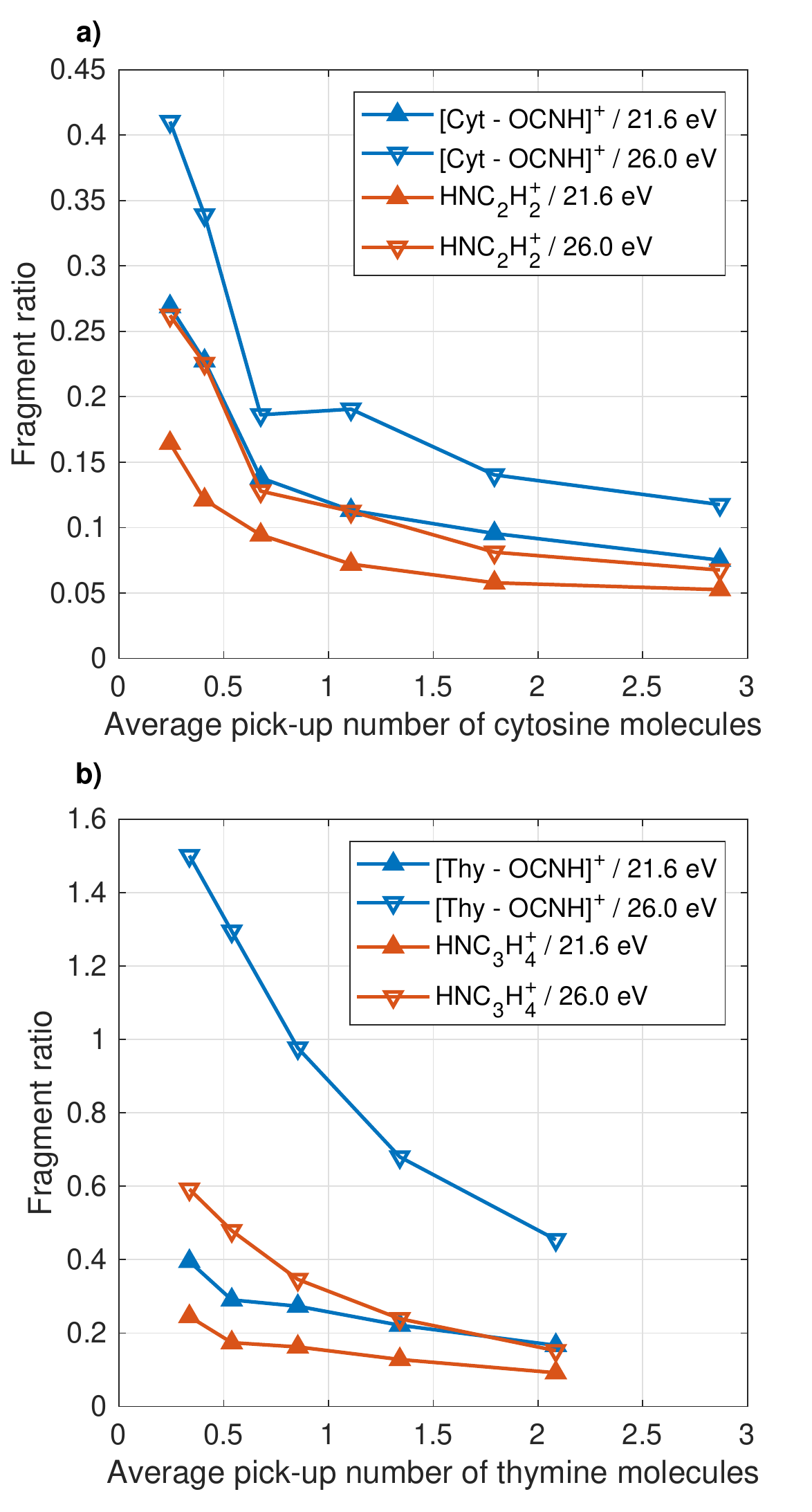}
    \caption{Ratios of the yield of the two main fragments of Cyt (a) and Thy (b) to the parent ion monomer and cluster ($n=1$-$4$) for increasing doping level of Cyt/Thy. These data were obtained for Penning ionization ($h\nu = 21.6$~eV) and charge-transfer ionization ($h\nu = 26$~eV).   }
    \label{fig:frag_doped}
\end{figure}
Reduced fragmentation of glycine and tryptophan from RCT ionization in HNDs has been observed for increased level of co-doping level with water.~\cite{ren2007changing,ren2008suppressing}
While microhydration clearly suppresses fragmentation of Cyt and Thy, naturally the question arises to to what extend this ``buffer effect'' is a unique property of water. To answer this question, we performed  comparative measurements where homogeneous Cyt/Thy clusters where formed in the HNDs. We measured the relative yield of the fragments to the parent ion and parent cluster ion for increased pure Cyt/Thy dopant clusters sizes. 
Fig.~\ref{fig:frag_doped} shows the ratio of the yields of the two main fragments to total yield of the parent ions and parent ion clusters (Cyt$_{1-4}^+$, Thy$_{1-4}^+$) as a function of average dopant cluster size. The relative reduction in the dopant fragment ratio is 60-70\,\% when increasing the average number of doped molecules from 0.5 to 2 in both cases of Penning and RCT ionization. 
A similar reduction of fragment yields with respect to monomers was observed for the case of impact ionization by keV ions.~\cite{schlatholter2006ion} New ion fragments were found for increasing Thy cluster sizes in a molecular beam which were facilitated by the intermolecular hydrogen-bonds in the cluster. These cluster-specific fragmentation channels are not present in HNDs most likely because conformations corresponding to local energy minima tend to be frozen out in HNDs at the expense of the equilibrium structures.~\cite{davies2019dimers,mani2019accessing}   

The fact that the relative fragment yield decreases so rapidly for increasing pyrimidine cluster size without co-doping water indicates that the reduced fragmentation is a general trend for increasing cluster size largely independent of the molecular composition. 
In the case of glycine and tryptophan in HNDs~\cite{ren2007changing,ren2008suppressing}, the degree to which fragmentation was buffered by water was different indicating dependencies on the intermolecular bonding in the dopant cluster. 
Thus, the buffering effect of fragmentation may be different for the purine bases (adenine and guanine) in HNDs and could possibly out-compete the buffering effect from formation of homogeneous purine clusters in the droplet. 
A systematic study of different combinations of co-dopants combined with quantum chemical calculations should be carried out to unravel the specific intermolecular effect of clusters on the suppression of fragmentation upon ionization. 


\begin{figure}[]
    \centering
    \includegraphics[width=0.9\columnwidth]{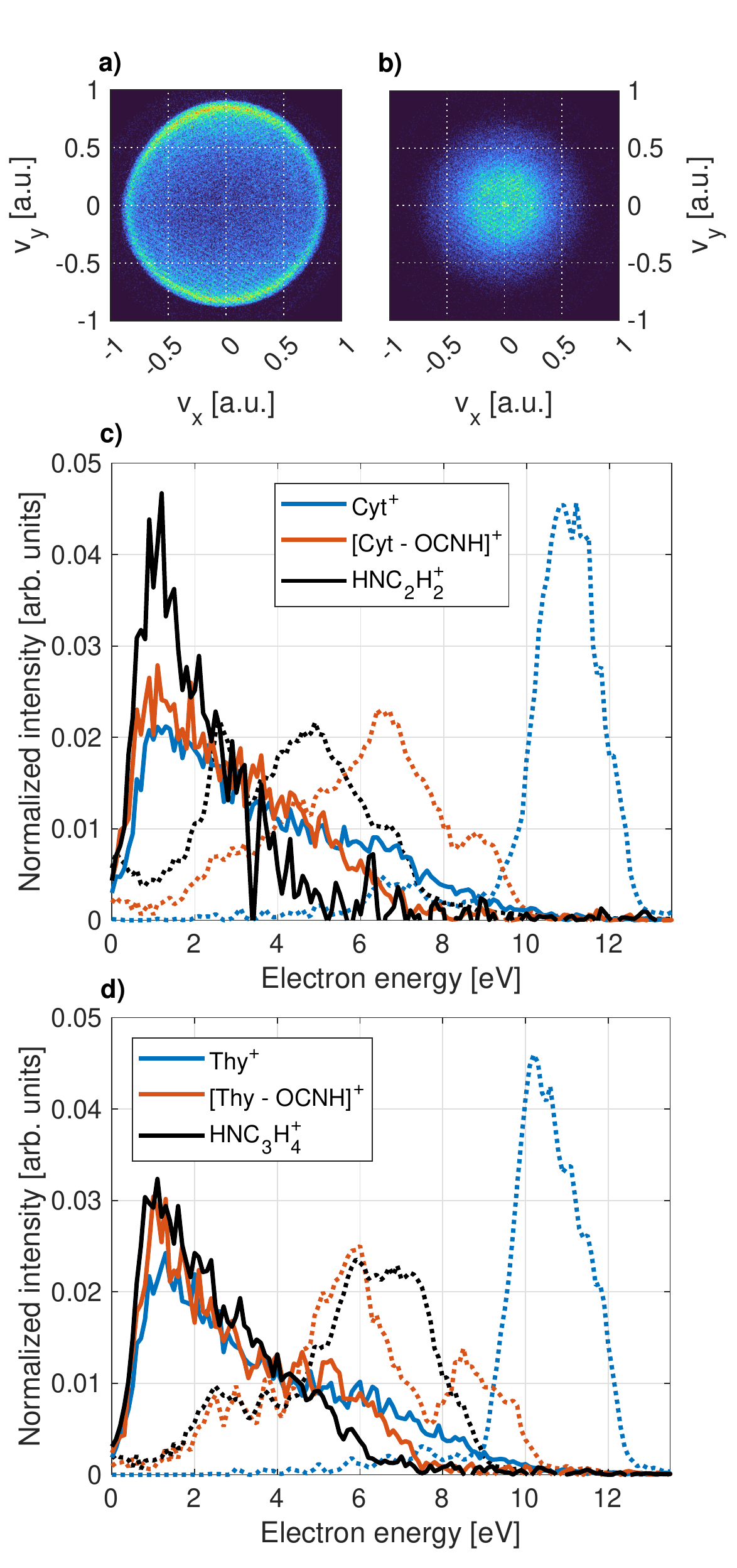}
    \caption{VMIs of electrons in coincidence with Cyt$^+$ recorded for (a) direct photoionization of effusive Cyt ($h\nu=20.6$~eV)  and (b) Penning ionization of Cyt in HNDs ($h\nu=21.6$~eV). Panel (c) and (d) show PIES ($h\nu=21.6$~eV) of HND-doped Cyt and Thy, respectively, showing the three main electron-ion coincidence channels (solid lines). The dotted lines show the corresponding electron spectra of effusive Cyt and Thy ($h\nu=20.6$~eV). }
    \label{fig:penning_doped}
\end{figure}

\begin{figure}[t]
    \centering
    \includegraphics[width=0.8\columnwidth]{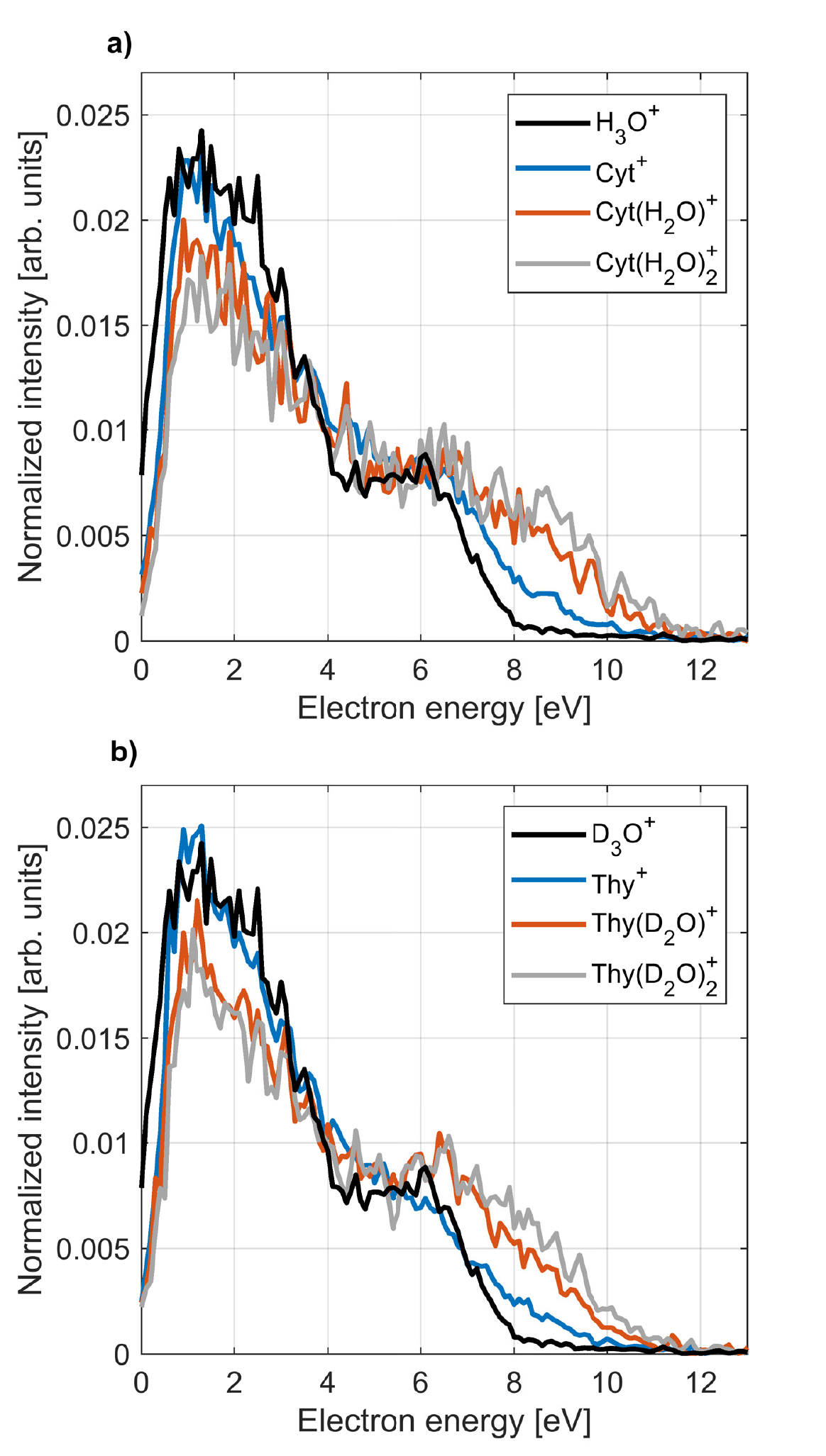}
    \caption{PIES ($h\nu=21.6$~eV) of microsolvated Cyt (a) and Thy (b) in coincidence with the main water cluster ion fragment (H$_3$O$^+$/D$_3$O$^+$), the nucleobase parent ion and the ionic cluster of the parent ion bound to one or two water molecules.  }
    \label{fig:penning_codoped}
\end{figure}

\subsection{Electron spectra}
The PIES of dopants in HNDs are notoriously broad and structure-less.~\cite{shcherbinin2018penning,mandal2020penning,ltaief2021photoelectron} An exception to this is are alkali metals which reside on the droplet surface.~\cite{buchta2013charge,ben2019charge,asmussen2022time} Fig.~\ref{fig:penning_doped} shows PIES recorded at $h\nu = 21.6$~eV and the corresponding PES recorded for an effusive beam at $h\nu = 20.6$~eV in coincidence with the two main fragments and the parent ion for Cyt and Thy. The electron spectra recorded for effusive Cyt and Thy are consistent with those previously reported given the limited resolution of our VMI spectrometer.~\cite{jochims2005photoion,trofimov2005photoelectron} The energy gap between the falling edges of the electron spectra for different coincidences towards larger kinetic energy matches the appearance energy of the different fragments reported for Thy.~\cite{jochims2005photoion} 
In contrast, PIES of Cyt/Thy in HNDs are broad and structureless resembling previously reported PIES of polyatomic molecules embedded in HNDs. The nature of the PIES in HNDs is still not well-described. We have previously modelled the PIES of acene molecules in HNDs by relaxation through a series of elastic electron-He binary collisions.~\cite{shcherbinin2018penning} However, the droplet size had to be massively overestimated to reach similar energy loss as observed in the experiment. 
We have recently demonstrated that the electron-He binary collision model can indeed reproduce the energy loss and change in angular distribution in the case of direct photoemission from HNDs.~\cite{asmussen2023electron} These two results imply that elastic scattering cannot alone account for the observed PIES. The sharp drop of signal at $<1$~eV was explained as electrons being trapped in the HNDs due to the barrier potential to the conduction band in liquid He which the electrons have to overcome to escape from the droplet.~\cite{rosenblit1995dynamics,asaf1986energies,buchenau1991excitation,wang2008photoelectron}
The falling edge of the PIES of acene molecules in HNDs could be modeled by a Maxwell-Boltzmann distribution corresponding to a thermal electron distribution of $>10^4$~K. However, the temperature is incompatible with the temperature of HNDs (0.4 K).~\cite{toennies2004superfluid} 
We note that such hot thermal emission of electrons has been described as ``hot electron ionization'' by Hansen \textit{et al.}~\cite{hansen2018hot} for single photon ionization~\cite{hansen2017single,andersson2023single}, multi-photon ionization~\cite{campbell2000above,kjellberg2010momentum} and Penning ionization~\cite{weber1998penning} of fullerenes. If hot electron ionization would apply to Penning ionization of Cyt/Thy in HNDs, it begs the question why hot electron ionization does not apply to direct ionization of the molecules in an effusive beam. Possibly, the presence of Cyt/Thy clusters in the droplet could promote for a different ionization mechanism; however at the doping conditions in the experiment mostly monomer doping condition is expected. Thus, there is little evidence available at the moment to determine if the PIES of Cyt/Thy is caused by hot electron ionization.  
PIES of rare gas atoms~\cite{wang2008photoelectron}, small organic molecules~\cite{mandal2020penning} and water (see Fig.~\ref{fig:penning_codoped}) in HNDs show additional features that can be associated with direct ionization from interatomic decay with an excited He atom. Additional experimental and theoretical work is needed to unravel the dopant-specific structure of the Penning ionization process in HNDs. 

Fig.~\ref{fig:penning_codoped} shows the PIES recorded for microsolvated Cyt and Thy. The PIES recorded in coincidence with Cyt$^+$ or Thy$^+$ is identical in the case of doping only with the nucleobases and co-doping with water whereas the PIES detected in coincidence with Cyt(H$_2$O)$_{1-2}^+$ and Thy(D$_2$O)$_{1-2}^+$ extends to higher kinetic energy. 
This suggests that fragmentation of the microsolvated complex to the bare nucleobase ion is weak. 
Thus, the yields of Cyt$^+$ and Thy$^+$ reflect Penning ionization in the case where no water molecules were captured by the droplet.  

We note that the Penning ionization electron distribution extends to higher kinetic energy for the ion fragment associated to the smallest appearance energy. The difference in highest detected electron energy for the different fragments matches the difference in appearance energy which indicates that the Penning spectra can be associated to the appearance energy for a specific fragment of the dopant in HNDs. 
With this information, we can asses the PIES of microsolvated Cyt and Thy in Fig.~\ref{fig:penning_codoped}. 
The PIES detected in coincidence with Cyt(H$_2$O)$_{1-2}^+$ and Thy(D$_2$O)$_{1-2}^+$ extend to higher kinetic energy than the PIES of Cyt$^+$, Thy$^+$ or H$_3$O$^+$, D$_3$O$^+$. 
Photoionization of hydrated nucleobases in a molecular beam expansion have revealed a red-shift of $\sim$0.3~eV in the appearance energy for M(H$_2$O)$_{1-2}^+$~\cite{belau2007vacuum} which matches well with the observation of a more extended PIES in HNDs. The  Cyt(H$_2$O)$_{1-2}^+$/Thy(D$_2$O)$_{1-2}^+$ shows a maximum in the PIES at $\sim$7~eV which is close to the maximum found in the PIES of H$_3$O$^+$, D$_3$O$^+$ ($\sim6$~eV). This indicates that the PIES of microhydrated dopants can be correlated to the PIES of water. This is expected, as the water molecules doped into the HNDs after the Cyt/Thy molecules form a hydration shell and Penning ionization is particularly surface-sensitive. Thus, despite their broad structure, the PIES give some insight into the secondary ionization of hydrated systems. However, more detailed understanding regarding the nature of the PIES of dopants is needed for any further analysis of the system.

\section{Conclusion}
In summary, we have presented mass spectra of Cyt and Thy and their microhydrated derivatives following Penning ionization and charge-transfer ionization in HNDs upon EUV irradiation. Fragmentation of the pyrimidine nucleobases is strongly suppressed upon ionization in helium nanodroplets compared to direct photoionization of the molecules in an effusive beam. Generally, the probability for the parent ion to fragment is smaller for Penning ionization making it a softer ionization process as compared to charge-transfer ionization. By increasing the dopant cluster size, either of the pure pyrimidine bases or by increasing the level of hydration of the pyrimidine molecules, fragmentation of the parent ion is also reduced for both ionization channels. The mass spectra of ionized Cyt and Thy clusters and microhydrated Cyt/Thy clusters in helium droplets deviate from clusters formed in a molecular beam expansion. The difference can most likely be assigned to the cold (0.4~K) environment of the helium droplets which facilitates stabilization of local minimum-energy structures. 

Penning ionization electron spectra of Cyt/Thy and their microhydrated derivatives are broad and dissimilar to photoelectron spectra of effusive Cyt/Thy. 
Nevertheless, the highest kinetic energy of the Penning ionization electron spectra recorded in coincidence with ion fragments reflect the appearance energy of that fragment for direct photoionization. 

These results demonstrate how a model system for radiation damage, \textit{i.~e.} microsolvated pyrimidine nucleobases, can be formed and studied in HNDs. Due to the efficient cooling of embedded species, helium droplets can give access to conformations not achievable under conventional molecular beams conditions. To obtain a more detailed understanding of the local minimum-energy conformations formed in HNDs and the quenching of fragmentation, further experiments and quantum chemistry calculations should be carried out.

\section*{Author Contributions}
J.D.A., A.R.A., A.S., B.B., K.S., S.D., L.B.L and M.M. performed the experiments with support from H.B.P.. S.K. aided remotely in the interpretation of the experimental results. J.D.A. and M.M. wrote the manuscript with input from all the co-authors.

\section*{Conflicts of interest}
There are no conflicts to declare.

\section*{Acknowledgements}
J.D.A. and M.M. acknowledge financial support by the Carlsberg Foundation. 
A.R.A. acknowledges with gratitude for the support from the Marie Skłodowska-Curie Postdoctoral Fellowship project Photochem-RS-RP (Grant Agreement No. 101068805) provided by the European Union’s Horizon 2020 Research and Innovation Programme.
S.R.K. thanks Dept. of Science and Technology, Govt. of India, for support through the DST-DAAD scheme and Science and Eng. Research Board.
S.R.K., K.S. and S.D. acknowledge the support of the Scheme for Promotion of Academic Research Collaboration, Min. of Edu., Govt. of India, and the Institute of Excellence programme at IIT-Madras via the Quantum Center for Diamond and Emergent Materials. 
S.R.K. gratefully acknowledges support of the Max Planck Society's Partner group programme. 
S.R.K acknowledges support for this research through the Indo-French Center for Promotion of Advanced Research (CEFIPRA)
M.M. and S.R.K. gratefully acknowledge funding from the SPARC Programme, MHRD, India. 
L.B.L. and M.M. acknowledge financial support by the Danish Council for Independent Research Fund (DFF) via Grant No. 1026-00299B.
The research leading to this result has been supported by the COST Action CA21101 ``Confined Molecular Systems: From a New Generation of Materials to the Stars (COSY)''.



\balance

\renewcommand\refname{References}

\bibliography{rsc} 
\bibliographystyle{rsc} 

\end{document}